# Microfluidic tools for assaying immune cell function

*Joel Voldman - Massachusetts Institute of Technology*

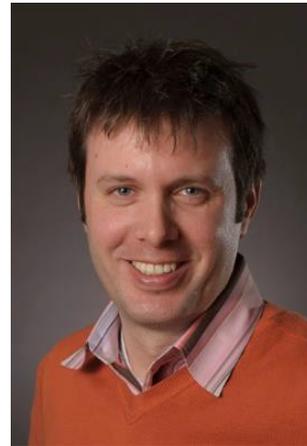

## Biography

*Joel Voldman is a Professor in the Electrical Engineering and Computer Science Department at MIT. He received the B.S. degree in electrical engineering from the University of Massachusetts, Amherst, in 1995. He received the M.S and Ph.D. degree in electrical engineering from the Massachusetts Institute of Technology (MIT), Cambridge, in 1997 and 2001, developing bioMEMS for single-cell analysis. Following this, he was a postdoctoral associate in George Church's lab at Harvard Medical School, where he studied developmental biology. In 2002 he returned to MIT as an Assistant Professor in the Electrical Engineering and Computer Science department at MIT. In 2004 he was awarded the NBX Career Development Chair, in 2006 promoted to Associate Professor, and in 2013 promoted to Professor in the department. Among several awards, he has received an NSF CAREER award, an ACS Young Innovator Award, a Bose Fellow award, Jamieson Teaching Award, Smullin Teaching Award, Quick Faculty Research Innovation Fellowship, and awards for posters and presentations at international conferences.*

*Prof. Voldman's research focuses on developing microfluidic technology for biology and medicine, with an emphasis on cell sorting and stem cell biology. He has developed a host of technologies to arrange, culture, and sort diverse cell types including immune cells, endothelial cells, and stem cells. Current areas of research include recapitulating the induction of atherosclerosis on a microfluidic chip, and using microfluidic tools to study how immune cells decide to attack tumor cells. He is also interested in translational medical work, such as developing point-of-care drop-of-blood assays for proteins and rapid microfluidic tests for immune cell activation for the treatment of sepsis.*

## Abstract

Microsystems have the potential to impact biology by providing new ways to manipulate cells and the microenvironment around them. Simply physically manipulating cells or their environment—using microfluidics, electric fields, or optical forces—provides new ways to separate cells and organize cell- cell interactions. Immune cells are of particular interest because of their central role in defending the body against foreign invaders. As a consequence, many microfluidic devices have been used to study both the basic biology of immune cells as well as to assay them for clinicaluse.

Our lab has developed technologies on both ends of the spectrum, from cell pairing devices able to study information flow in immune cells, to electrical sorting devices for assaying immune cell function in response to disease.

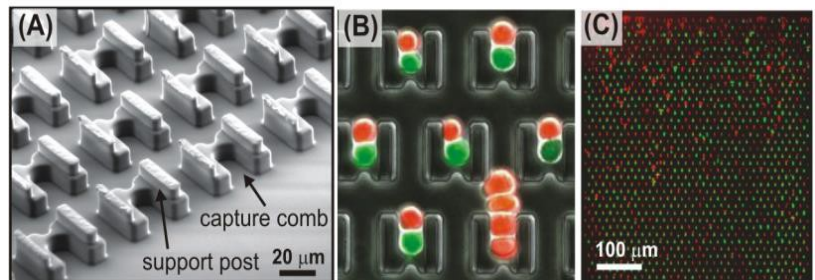

*Figure 1. High-throughput cell pairing and fusion. (A) Device overview, (B) close-up of cell pairing, (C) Pairing over the entire array.*

In terms of cell pairing, we have developed two complementary approaches to creating programmed pairs of cells, one using capture "cups" and a three-step back-and-forth loading procedure to pair thousands of cells in parallel[1,2], and the other using microfluidic "corrals" to contain cells[3,4] (Figure 1). With these devices we can pair immune cells with each other or with other cells (i.e., tumor cells to study information flow from first contact to downstream effector functions, elucidating how decision-making occurs in these interactions.

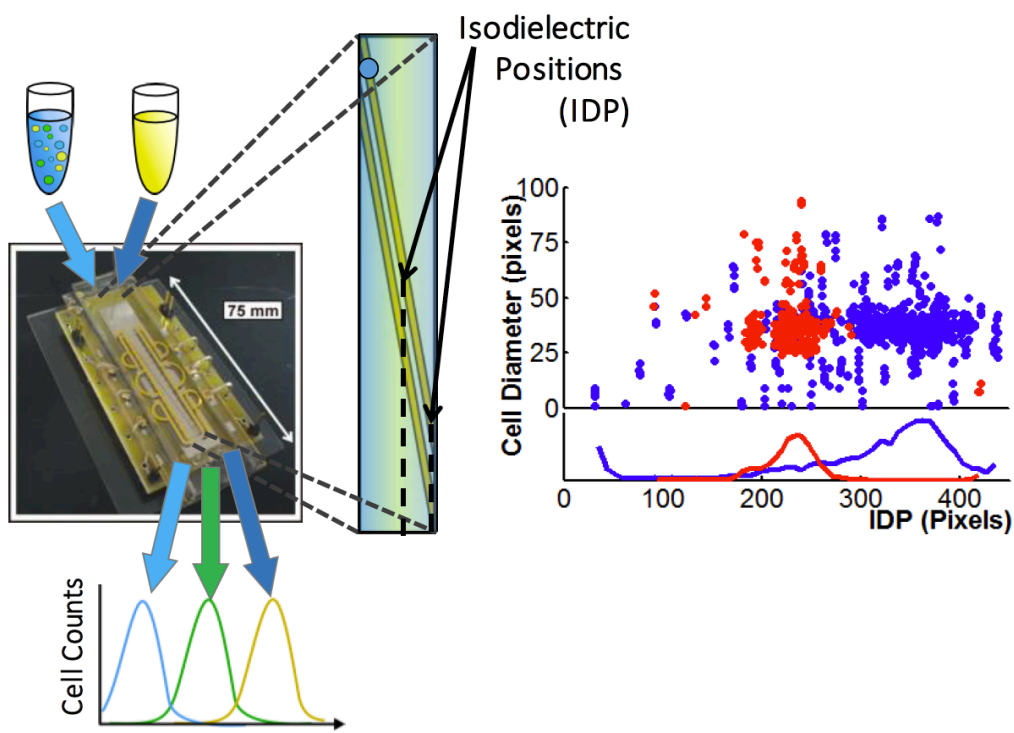

*Figure 2. Iso-dielectric separation. A heterogeneous cell population (blue tube) is introduced in a microfluidic device where the cells encounter a spatial gradient in liquid conductivity and a dielectrophoretic force that pushes them across this gradient, until they reach their iso-dielectric point (IDP), where the force goes to zero and the cells cross over. At right are histograms of cell IDPs for unactivated (blue) and activated (red) human neutrophils.*

In terms of electrical sorting devices, we have developed microfluidic systems to sort cells based on their intrinsic electrical properties. Electrical properties have previously been correlated with important biological phenotypes (apoptosis, cancer, etc.), but a sensitive and specific method approach has been lacking. We have developed a method called iso-dielectric separation that uses electric fields to drive cells to the point in a conductivity gradient where they become electrically transparent, resulting in a continuous separation method specific to electrical properties[5-8]. With this method, we have screened the entire genome of an organism to understand the biological basis of electrical properties, finding that the relationship between genetics and intrinsic properties has both intuitive and non-intuitive features.

## References

1. Skelley, A.M., Kirak, O., Suh, H., Jaenisch, R. & Voldman, J. Microfluidic Control of Cell Pairing and Fusion. *Nature Methods* **6**, 147-152 (2009). PMC:PMC3251011.
2. Dura, B., Dougan, S.K., Barisa, M., Hoehl, M.M., Lo, C.T., Ploegh, H.L. & Voldman, J. Profiling Lymphocyte Interactions at the Single-Cell Level by Microfluidic Cell Pairing. *Nat Commun* **6**, 5940 (2015).
3. Dura, B., Servos, M.M., Barry, R.M., Ploegh, H.L., Dougan, S.K. & Voldman, J. Longitudinal Multiparameter Assay of Lymphocyte Interactions from Onset by Microfluidic Cell Pairing and Culture. *Proc Natl Acad Sci U S A* **113**, E3599-3608 (2016).
4. Dura, B., Liu, Y. & Voldman, J. Deformability-Based Microfluidic Cell Pairing and Fusion. *Lab on a Chip* **14**, 2783-2790 (2014).
5. Vahey, M.D. & Voldman, J. An Equilibrium Method for Continuous-Flow Cell Sorting Using Dielectrophoresis. *Analytical Chemistry* **80**, 3135-3143 (2008).
6. Vahey, M.D. & Voldman, J. High-Throuput Cell and Particle Characterization Using Isodielectric Separation. *Analytical Chemistry* **81**, 2446-2455 (2009). PMC:PMC2675787.
7. Vahey, M.D., Pesudo, L.Q., Svensson, J.P., Samson, L.D. & Voldman, J. Microfluidic Genome-Wide Profiling of Intrinsic Electrical Properties in Saccharomyces Cerevisiae. *Lab on a Chip* **13**, 2754-2763 (2013). PMC:PMC3686985.
8. Prieto, J.L., Su, H.-W., Hou, H.W., Vera, M.P., Levy, B.D., Baron, R.M., Han, J. & Voldman, J. Monitoring Sepsis Using Electrical Cell Profiling. *Lab on a Chip* (2016).